\documentclass[11pt,a4paper]{emulateapj}
\usepackage{verbatim}
\usepackage{epsfig}
\usepackage{amsmath}
\usepackage{natbib}

\begin{document}
 
\title{A Ground-Based Albedo Upper Limit for HD 189733\lowercase{b} from Polarimetry}

\author{Sloane J. Wiktorowicz}
\affil{Department of Astronomy and Astrophysics, University of California, Santa Cruz, CA 95064}

\author{Larissa A. Nofi}
\affil{Department of Astronomy and Astrophysics, University of California, Santa Cruz, CA 95064}
\affil{Institute for Astronomy, University of Hawaii, Honolulu, HI 96822}

\author{Daniel Jontof-Hutter}
\affil{NASA Ames Research Center, Moffett Field, CA 94035}
\affil{Department of Astronomy, Davey Laboratory, Pennsylvania State University, University Park, PA 16802}

\author{Pushkar Kopparla}
\affil{Division of Geological and Planetary Sciences, California Institute of Technology, Pasadena, CA 91125}

\author{Gregory P. Laughlin, Ninos Hermis}
\affil{Department of Astronomy and Astrophysics, University of California, Santa Cruz, CA 95064}

\author{Yuk L. Yung}
\affil{Division of Geological and Planetary Sciences, California Institute of Technology, Pasadena, CA 91125}

\and

\author{Mark R. Swain}
\affil{Jet Propulsion Laboratory, California Institute of Technology, 4800 Oak Grove Drive, Pasadena, CA 91109}

\email{sloanew@ucolick.org}

\date{\today}

\begin{abstract}

We present 50 nights of polarimetric observations of HD 189733 in $B$ band using the POLISH2 aperture-integrated polarimeter at the Lick Observatory Shane 3-m telescope. This instrument, commissioned in 2011, is designed to search for Rayleigh scattering from short-period exoplanets due to the polarized nature of scattered light. Since these planets are spatially unresolvable from their host stars, the relative contribution of the planet-to-total system polarization is expected to vary with an amplitude of order 10 parts per million (ppm) over the course of the orbit. Non-zero and also variable at the 10 ppm level, the inherent polarization of the Lick 3-m telescope limits the accuracy of our measurements and currently inhibits conclusive detection of scattered light from this exoplanet. However, the amplitude of observed variability conservatively sets a $3 \sigma$ upper limit to the planet-induced polarization of the system of 58 ppm in $B$ band, which is consistent with a previous upper limit from the POLISH instrument at the Palomar Observatory 5-m telescope \citep{Wiktorowicz2009}. A physically-motivated Rayleigh scattering model, which includes the depolarizing effects of multiple scattering, is used to conservatively set a $3 \sigma$ upper limit to the geometric albedo of HD 189733b of $A_g < 0.37$. This value is consistent with the value $A_g = 0.226 \pm 0.091$ derived from occultation observations with \textit{HST STIS} \citep{Evans2013}, but it is inconsistent with the large $A_g = 0.61 \pm 0.12$ albedo reported by \cite{Berdyugina2011}. \\

\end{abstract}

\section{Introduction}

Since the polarization of incident starlight scattered by a planetary atmosphere depends on the morphology, size, index of refraction, and vertical distribution of the scattering particles, scattered light polarimetry of planets presents a rich opportunity for the study of their atmospheres. In the Solar System, polarimetry has provided fascinating results for both Venus and Titan. For Venus, the significant negative branch polarization (i.e., with polarization vector oriented parallel to the Sun-Venus-observer ``scattering plane"), and the peculiar variation of polarization as a function of phase angle \citep{Lyot1929, Coffeen1969}, are consistent with spherical, $1.05 \pm 0.10$ $\mu$m radius cloud particles composed of a concentrated sulfuric acid solution \citep{Hansen1974}. In contrast, Rayleigh scattering imparts positive branch polarization, where polarization is oriented perpendicular to the scattering plane. \\

For Titan, large photochemical haze particles are suggested by the intensity in forward scattering \citep{Rages1983}, but small particles are implied by Titan's strong polarization ($P \sim 50\%$) at $\sim 90^\circ$ scattering angles from Pioneer 11 \citep{Tomasko1982} and Voyager 1 and 2 observations \citep{West1983}. \cite{West1991} suggested fractal aggregates for the shape of the aerosols, and \cite{WestSmith1991} showed that such particles could reconcile the measurements of high polarization and strong forward scattering. Thus, the combination of polarimetry and photometry enabled the discovery that Titan's large, fractal haze particles are composed of thousands of small, spherical monomers \citep{Tomasko1982, West1983, Tomasko2009}. Unfortunately, since polarimetry is most powerful when studying scattering through a large range in phase angles, the utility of ground-based polarimetry for most Solar System objects is limited. However, such a limitation is not present for most exoplanets, because time-variable phase angle $\alpha(t)$ is given by $\cos[ \alpha(t) ] = \sin{i} \cos [2 \pi (t / T - 0.5) ]$ for orbital inclination $i$ and period $T$ on a circular orbit (where $t = 0$ indicates mid-transit or inferior conjunction of the planet). Given the expectation value for randomly distributed orbital inclinations, $i_{\rm exp} \sim 52^\circ$, most exoplanets will traverse between $38^\circ < \alpha < 142^\circ$. Therefore, short-period exoplanets not only quickly sweep through a large range in phase angles, pronouncing them as desirable targets for scattered light polarimetry, but they also maximize intercepted starlight. \\

The hot Jupiter HD 189733b is an intriguing target for this study because of its large radius and close orbit, which maximize relative photon counts, and the brightness of its host star, which maximizes absolute photon counts. Interestingly, while a haze of small, Rayleigh-scattering particles is interpreted to be present in the atmosphere from \textit{HST STIS}, \textit{ACS}, and \textit{WFC3} observations \citep{Lecavelier2008, Pont2008, Sing2009, Sing2011, Gibson2012, Huitson2012, Pont2013}, this may be supplanted by the inaccurate subtraction of unocculted starspots \citep{McCullough2014}. While starspots may induce symmetry breaking of the stellar limb polarization \citep{Chandrasekhar1946a, Chandrasekhar1946b}, this effect is modeled to be at the ppm level or below in linear polarimetry, because the cross-sectional area of a starspot vanishes at the limb \citep{Berdyugina2011}. However, transiting exoplanets are expected to impart symmetry breaking polarization at ingress and egress at the 10 ppm level or below in broadband \citep{Carciofi2005, Kostogryz2011, WiktorowiczLaughlin2014, Kostogryz2015}. Investigation of this effect is beyond the scope of this paper. As with Titan, the presence of haze particles in HD 189733b may best be tested with a combination of photometry and polarimetry. \\

Surrounding occultation, when the exoplanet's disk dives behind the limb of the host star, \cite{Evans2013} find a significant change in the brightness of the system with \textit{HST STIS}, which is interpreted as scattered light from the planet. Indeed, a weighted mean of the inferred geometric albedos of the planet in the $390-435$ nm and $435-480$ nm channels suggests a $B$ band albedo of $A_g = 0.226 \pm 0.091$. However, variability in the $B$ band linear polarization of the system was reported with an amplitude of order 100 ppm, which suggests a geometric albedo of $A_g = 0.61 \pm 0.12$ \citep{Berdyugina2008, Berdyugina2011}. Our original polarimetric investigation found a $99\%$ confidence upper limit to the variability of the system of 79 ppm, but these observations were taken unfiltered in a broader and redder bandpass \citep{Wiktorowicz2009}. In an updated analysis of the sensitivity of the measurement, taking into account instrumental modulation efficiency, we determine the bandpass of the \cite{Wiktorowicz2009} investigation to be $320-633$ nm with a central wavelength of 437 nm. Regardless, both photometric and polarimetric investigations have provided inconclusive evidence for Rayleigh scattering in the atmosphere of HD 189733b. \\

We seek to utilize our significantly expanded dataset to constrain the polarimetric amplitude, and therefore the albedo of the Rayleigh scattering surface, for this exoplanet. We briefly discuss the POLISH2 polarimeter at the Lick Observatory Shane 3-m telescope in Section \ref{sec_polish} and observations of the HD 189733 system in Section \ref{sec_results}. In Section \ref{sec_discuss}, we discuss our observations in the context of this nascent field as well as paths toward improvements in data quality. Finally, we present concluding remarks in Section \ref{section_conclusion}. \\

\section{Methods}
\label{sec_polish}

\subsection{The POLISH2 Polarimeter}

In this section, we briefly describe POLISH2 (POlarimeter at Lick for Inclination Studies of Hot jupiters 2), and we refer the reader to \cite{WiktorowiczNofi2015} for further inquiry. Rather than using a conventional half waveplate to convert incident polarization into an intensity modulation that may be measured by conventional imaging detectors, POLISH2 employs two photoelastic modulators (PEMs). The stress birefringent property of the fused silica PEMs preferentially retards the electric field component oriented $\pm 45^\circ$ from the stress axis. When coupled with a Wollaston prism, the PEMs impart a nearly sinusoidal intensity modulation on the photomultiplier tube detectors. The resonant frequencies of the PEMs are 40 and 50 kHz, which causes linear and circular polarization to modulate at linear combinations of these frequencies. After high speed digitization of the detector outputs, software demodulation simultaneously measures Stokes $q = Q/I$, $u = U/I$ (fractional linear polarization), and $v = V/I$ (fractional circular polarization), which describe the fractional polarization of incident light. \\

While spatially resolved circular polarization of Jupiter has been detected at the $\sim100$ ppm level from multiple scattering \citep{Kemp1971a, Kemp1971b, Michalsky1974}, the sign of circular polarization is observed to reverse between northern and southern hemispheres. For exoplanets, it is expected that the dilution of circular polarization by direct light from the host star, as well as from integrating over the planetary disk, will cause exoplanet circular polarization to be more difficult to measure than linear polarization. \\

The two major improvements in the use of PEMs over waveplates are as follows: 1) simultaneous Stokes $q$ and $u$ measurements obviate systematic effects from waveplate rotation (heterogeneity in retardance across the optic itself), from atmospheric or astrophysical changes on short timescales, and potentially doubles the throughput of the measurement; and 2) high speed modulation enables photon-limited sensitivity via $1 / \sqrt{n_\text{modulations}}$, where ``modulation" is defined by sequential measurement of Stokes $\pm q$, for example. In contrast to PEMs, with a modulation timescale of order 10 $\mu$s, typical waveplate modulations have a duration of order minutes to minimize overhead due to waveplate rotation. \\

\subsection{Observations and Calibration}
\label{sec_obs}

To directly detect scattered light from spatially unresolvable exoplanets, the ability to measure nightly changes in the polarization of starlight at the 10 ppm level or below is required. In this regime, accuracy (the ability to calibrate non-astrophysical change to some level) is far more important than sensitivity (the ability to measure a change to that level). While photon noise defines the fundamental limit to the accuracy of a measurement, instrumental and atmospheric systematic effects tend to dominate for many exoplanet investigations. Therefore, it is crucial that these systematic effects be removed or calibrated to the 1 ppm level to enable confident detection of scattered light from exoplanets at the 10 ppm level. We identify and correct for two dominant systematics: polarized sky foreground and non-zero telescope polarization. \\

As the POLISH2 field of view is $\sim 15\arcsec$ in diameter, a detectable quantity of sky photons is present even in $B$ band. This foreground tends to be polarized, especially when the Moon lies $\sim 90^\circ$ from the target. To mitigate this systematic effect, we perform one, 30-sec integration on a sky field $30\arcsec$ N of the target for every two integrations on-target. Therefore, one-third of all telescope time is devoted to monitoring sky polarization with a cadence of one minute. \\

At the focus of a telescope, detectable polarization is measured even when observing an unpolarized star. This is because reflectivity variations across the telescope mirrors cause a discrepancy in the intensities of pairs of light rays with equal but opposite incidence angles, which causes the polarization from certain patches of the mirror to dominate. Even at Cassegrain focus of a variety of telescopes, this so-called ``telescope polarization" is found by many authors to be 10 to 100 ppm in amplitude \citep{Hough2006, Wiktorowicz2008, Lucas2009, Wiktorowicz2009, Berdyugina2011, WiktorowiczNofi2015}. \\


The standard approach for measuring telescope polarization is to observe a handful of nearly unpolarized calibrator stars. In practice, however, the details of this process may impart additional systematics that may be misinterpreted as arising from the science target. For instance, \cite{Lucas2009} observe that 12 of their 75 telescope polarization measurements are inconsistent with the mean value of telescope polarization at the $> 3\sigma$ level, and they consequently reject these measurements. However, the probability of this occurrence in a normally distributed population is $\sim 10^{-9}$. \cite{Lucas2009} therefore note that the measurements must have a non-Gaussian distribution and advocate for repeated measurements if possible. We concur with this request, and we also caution that rejecting calibrator measurements that lie $> 3 \sigma$ from the mean will introduce a bias into the science results, because measurements may not be drawn from a Gaussian distribution. \\

In addition, we advocate that the observing cadence on calibrator stars follow that of the science targets, lest biases be introduced. \cite{Wiktorowicz2009} observe the same calibrator star during each night in the study, but the use of only one calibrator limits the accuracy of the zero point of polarization measurements. However, accurate measurement of the time-averaged polarization of an exoplanet host star is necessary only to describe the state of the intervening ISM dust particles and has no relevance to the scattering properties of an exoplanet atmosphere. \cite{Wiktorowicz2009} found the square root of the weighted variance of nightly telescope polarization observations to be $\sigma_q = 7.5$ ppm and $\sigma_u = 5.3$ ppm \citep[Table 4]{Wiktorowicz2009}, while the upper limit to exoplanet variation was found to be 79 ppm. Therefore, observation of calibrator and science targets with the same nightly cadence enabled calibrations to be an order of magnitude more accurate than science observations. Observing the same five calibrator stars nearly every night, \cite{Lucas2009} expand this approach to accurately measure telescope polarization each night. While \cite{Berdyugina2011} observe 26 calibrator stars over 15 nights, the dates of observation for each star are not presented. Indeed, the long duration of each calibrator star observation, one to two hours, implies that very few calibrators were observed on successive nights. In an investigation where systematic effects are severe, it is imperative that the same control and science targets be observed during each night. \\

Therefore, we observe four to eleven telescope polarization calibrator stars during each night of a given run, and the nightly target list is reproduced for each night of the run. Depending on the time of year, the makeup of this list will of course vary based on observability. These stars are identified from both the PlanetPol group \citep{Hough2006, Lucas2009, Bailey2010} and our own unpublished survey of nearby, bright stars with POLISH2 at the Lick Observatory Nickel 1-m telescope. Given that interstellar polarization scales roughly linearly with heliocentric distance (e.g., \citealp{Hall1949, Hiltner1949, Fosalba2002}), nearly unpolarized stars are expected to lie in the vicinity of the Sun. Due to this proximity, such suitable stars tend to be quite bright. Therefore, sufficient measurement sensitivity is typically obtained after nine minutes for each telescope polarization calibrator star. \\

Strongly polarized stars ($p = \sqrt{q^2 + u^2} \sim 1\%$) represent the second type of calibrator. These stars are typically observed to determine the rotational position of the instrument; that is, the relationship between instrumental Stokes $(q, u)$ and celestial $(q, u)$. We typically observe the same two strongly polarized stars for nine minutes during each night of a run. Finally, \cite{WiktorowiczNofi2015} discuss absolute calibration of POLISH2 via sequential, laboratory injection of 100\% Stokes $q$, $u$, and $v$. \\

\section{Results}
\label{sec_results}


We performed 50 nights of $B$ band observations of HD 189733 with POLISH2 at the Lick 3-m telescope between 2011 August 13 and 2014 July 14 UT. Observing duration typically varied from one to three hours per night, depending on the time of year and weather. Table \ref{stars} lists the stars observed during this program, where previous estimates of the degree of linear polarization are shown in the last column. Table \ref{189} presents nightly measurements of the linear and circular fractional polarization (Stokes $q$, $u$, and $v$) of the HD 189733 system subtracted by telescope polarization. These values are binned in orbital phase in Table \ref{189bin}, and they are subtracted by the time-averaged polarization listed in the first line of that table. Since the magnitude of a vector is biased in the presence of noise, we employ the generalized MAS estimator \citep{Plaszczynski2014} to estimate the degree of linear polarization $p$. \\

\begin{deluxetable*}{c c c c c r c c c c}
\tabletypesize{\normalsize}
\tablecaption{Observed Stars}
\tablewidth{0pt}
\tablehead{
\colhead{Star Name} & \colhead{HD} & \colhead{HR} & \colhead{RA (J2000)} & \colhead{Dec (J2000)} & \colhead{$d$ (pc)} & \colhead{$B$} & \colhead{Spec. Type} & \colhead{$P$ (ppm)} & \colhead{Ref.}}
\startdata
HD 189733	 & 189733		& $-$	& 20 00 43.7	& +22 42 39.06		& 19.5	& 8.58	 & K0V+M4V		& 300.7(6.4)	& 1 \\
\hline
$\alpha$ And 	 & 358		& 15	 	& 00 08 23.3	& +29 05 25.55		& 29.7	& 1.95	 & B8IV-VHgMn	& 100(1200)	& 2 \\
$\alpha$ Ari 	 & 12929		& 617	& 02 07 10.4	& +23 27 44.70		& 20.2	& 3.17	 & K1IIIb			& 300(1200)	& 2 \\
$\beta$ Tri 	 & 13161		& 622	& 02 09 32.6	& +34 59 14.27		& 38.9	& 3.14	 & A5III			& 500(1200)	& 2 \\
$\nu$ Tau 	 & 25490		& 1251	& 04 03 09.4	& +05 59 21.48		& 35.9	& 3.94	 & A0.5Va			& 400(1200)	& 2 \\
$\pi^3$ Ori 	 & 30652		& 1543	& 04 49 50.4	& +06 57 40.59		& 8.1		& 3.63	 & F6V			& 400(1200)	& 2 \\
38 Lyn 	 	 & 80081		& 3690	& 09 18 50.6	& +36 48 09.33		& 38.3	& 3.88	 & A1V			& 500(1200)	& 2 \\
$\theta$ UMa 	 & 82328		& 3775	& 09 32 51.4	& +51 40 38.28		& 13.5	& 3.64	 & F7V			& 100(1200)	& 2 \\
$\beta$ Leo 	 & 102647		& 4534	& 11 49 03.6	& +14 34 19.41		& 11.0	& 2.22	 & A3Va			& 2.3(1.1)		& 3 \\
$\beta$ Vir 	 & 102870		& 4540	& 11 50 41.7	& +01 45 52.99		& 10.9	& 4.15	 & F9V			& 3.3(1.4)		& 3 \\
$\eta$ Boo 	 & 121370		& 5235	& 13 54 41.1	& +18 23 51.79		& 11.4	& 3.25	 & G0IV			& 3.5(1.8)		& 3 \\
$\alpha$ Boo 	 & 124897		& 5340	& 14 15 39.7	& +19 10 56.67		& 11.3	& 1.18	 & K0III			& 6.3(1.6)		& 3 \\
$\gamma$ Boo & 127762		& 5435	& 14 32 04.7	& +38 18 29.70		& 26.6	& 3.21	 & A7III			& 3.6(1.6)		& 3 \\
$\zeta$ Her 	 & 150680		& 6212	& 16 41 17.2	& +31 36 09.79		& 10.7	& 3.43	 & G0IV			& 9.6(2.6)		& 3 \\
$\eta$ Her 	 & 150997		& 6220	& 16 42 53.8	& +38 55 20.11		& 33.3	& 4.42	 & G7.5IIIb			& 1300(1200)	& 2 \\
$\zeta$ Aql 	 & 177724		& 7235	& 19 05 24.6	& +13 51 48.52		& 25.5	& 3.00	 & A0IV-Vnn		& 22.8(3.0)	& 3 \\
$\iota$ Cyg 	 & 184006		& 7420	& 19 29 42.4	& +51 43 47.21		& 37.2	& 3.92	 & A5V			& $-$		& $-$ \\
$\alpha$ Aql 	 & 187642		& 7557	& 19 50 47.0	& +08 52 05.96		& 5.1	 	& 0.98	 & A7Vn			& 7.4(1.3)		& 3 \\
$\epsilon$ Cyg  & 197989	& 7949	& 20 46 12.7	& +33 58 12.93		& 22.3	& 3.52	 & K0III-IV			& 50(200)		& 2 \\
$\theta$ Peg 	 & 210418		& 8450	& 22 10 12.0	& +06 11 52.31		& 28.3	& 3.62	 & A1Va			& 500(1200)	& 2 \\
$\alpha$ Lac 	 & 213558		& 8585	& 22 31 17.5	& +50 16 56.97		& 31.5	& 3.78	 & A1V			& 400(200)	& 2 \\
\enddata
\label{stars}
\tablerefs{1. \cite{Wiktorowicz2009}, 2. \cite{Heiles2000}, 3. \cite{Bailey2010}}
\end{deluxetable*}

\begin{deluxetable*}{l l c c c c c}
\tabletypesize{\normalsize}
\tablecaption{Nightly HD 189733 Observations}
\tablewidth{0pt}
\tablehead{
\colhead{UT Date} & \colhead{MJD} & \colhead{$q$} & \colhead{$u$} & \colhead{$p$} & \colhead{$\Theta$} & \colhead{$v$} \\
& & (ppm) & (ppm) & (ppm) & ($^\circ$) & (ppm)}
\startdata
2011 Aug 13	& 55786.289(51)	& $+62(38)$	& $+120(32)$	& $131(33)$	& $31.4(7.9)$	& $-27(81)$ \\
2011 Aug 14	& 55787.286(54)	& $+34(31)$	& $+35(27)$	& $41(29)$	& $23(20)$	& $+69(68)$ \\
2011 Aug 15	& 55788.286(55)	& $+69(30)$	& $+33(26)$	& $72(30)$	& $13(11)$	& $+48(66)$ \\
2011 Aug 16	& 55789.287(55)	& $+46(31)$	& $+55(27)$	& $66(29)$	& $25(13)$	& $+40(69)$ \\
2011 Aug 17	& 55790.283(55)	& $+32(32)$	& $+81(27)$	& $82(28)$	& $34(11)$	& $+74(70)$ \\
2012 May 05	& 56052.455(37)	& $+98(33)$	& $-63(29)$	& $113(32)$	& $163.5(7.6)$	& $+149(73)$ \\
2012 May 06	& 56053.459(35)	& $+57(35)$	& $-101(30)$	& $110(32)$	& $149.7(8.8)$	& $-39(78)$ \\
2012 May 07	& 56054.448(41)	& $+18(30)$	& $+74(26)$	& $70(26)$	& $38(12)$	& $+136(67)$ \\
2012 May 08	& 56055.450(40)	& $+24(31)$	& $-6(26)$	& $16(30)$	& $173(51)$	& $-52(68)$ \\
2012 May 09	& 56056.446(42)	& $-59(29)$	& $+14(25)$	& $55(29)$	& $83(13)$	& $+104(65)$ \\
2012 Jun 07	& 56085.430(32)	& $+18(34)$	& $+47(30)$	& $40(30)$	& $35(22)$	& $+111(76)$ \\
2012 Jun 08	& 56086.432(30)	& $-5(36)$	& $+102(31)$	& $96(31)$	& $46(10)$	& $+52(79)$ \\
2012 Jun 09	& 56087.435(27)	& $-34(37)$	& $-59(32)$	& $59(33)$	& $120(17)$	& $-48(82)$ \\
2012 Jun 10	& 56088.456(17)	& $+10(47)$	& $+42(41)$	& $29(41)$	& $38(43)$	& $+60(100)$ \\
2012 Jun 11	& 56089.434(29)	& $+73(36)$	& $+5(30)$	& $67(36)$	& $2(13)$	& $+16(78)$ \\
2012 Jun 12	& 56090.436(28)	& $-42(36)$	& $+76(31)$	& $80(32)$	& $60(12)$	& $-54(79)$ \\
2013 May 24	& 56436.434(67)	& $+19(32)$	& $+169(27)$	& $166(27)$	& $41.9(5.5)$	& $+78(71)$ \\
2013 May 25	& 56437.428(71)	& $+15(31)$	& $+138(26)$	& $136(26)$	& $41.9(6.4)$	& $+51(67)$ \\
2013 May 26	& 56438.3569(91)	& $-56(81)$	& $+85(70)$	& $77(74)$	& $62(27)$	& $+140(180)$ \\
2013 May 27	& 56439.448(51)	& $-3(37)$	& $+82(33)$	& $73(33)$	& $46(14)$	& $+101(83)$ \\
2013 May 29	& 56441.410(58)	& $-40(46)$	& $+87(39)$	& $85(40)$	& $57(15)$	& $-71(100)$ \\
2013 May 30	& 56442.407(90)	& $-55(28)$	& $+143(24)$	& $150(25)$	& $55.5(5.3)$	& $-58(61)$ \\
2013 Aug 17	& 56521.307(73)	& $+48(40)$	& $+87(35)$	& $92(36)$	& $31(12)$	& $-70(90)$ \\
2013 Aug 18	& 56522.32(11)	& $-38(27)$	& $+98(22)$	& $102(23)$	& $55.6(7.3)$	& $-96(59)$ \\
2013 Aug 19	& 56523.34(10)	& $+22(29)$	& $+60(25)$	& $58(25)$	& $35(14)$	& $-23(63)$ \\
2013 Aug 20	& 56524.379(51)	& $+85(45)$	& $+58(38)$	& $95(43)$	& $17(12)$	& $-290(99)$ \\
2013 Aug 21	& 56525.31(11)	& $+107(27)$	& $+43(23)$	& $113(27)$	& $11.0(6.0)$	& $-236(61)$ \\
2013 Aug 22	& 56526.32(10)	& $+111(30)$	& $+41(25)$	& $116(29)$	& $10.1(6.3)$	& $-78(64)$ \\
2013 Sep 11	& 56546.29(10)	& $+18(26)$	& $+45(22)$	& $42(22)$	& $34(16)$	& $-55(56)$ \\
2013 Sep 12	& 56547.28(10)	& $-28(29)$	& $+22(25)$	& $28(27)$	& $71(27)$	& $-75(64)$ \\
2013 Sep 13	& 56548.279(99)	& $-51(25)$	& $+74(22)$	& $86(23)$	& $62.3(8.0)$	& $-63(57)$ \\
2013 Sep 14	& 56549.28(10)	& $-3(25)$	& $+32(21)$	& $24(21)$	& $48(27)$	& $-95(56)$ \\
2013 Sep 15	& 56550.275(99)	& $+58(24)$	& $+47(21)$	& $71(23)$	& $19.7(9.0)$	& $-75(54)$ \\
2013 Sep 16	& 56551.27(10)	& $+27(24)$	& $+47(20)$	& $49(21)$	& $30(13)$	& $-127(53)$ \\
2013 Oct 11	& 56576.237(95)	& $+8(26)$	& $+2(23)$	& $4(26)$		& $-$		& $-48(57)$ \\
2013 Oct 12	& 56577.228(80)	& $+68(28)$	& $+46(25)$	& $78(27)$	& $16.9(9.6)$	& $+166(65)$ \\
2013 Oct 13	& 56578.217(83)	& $-15(32)$	& $+3(27)$	& $9(32)$		& $-$		& $+68(69)$ \\
2013 Oct 14	& 56579.220(80)	& $-11(26)$	& $+2(23)$	& $7(26)$		& $-$		& $+15(58)$ \\
2014 Jun 07	& 56815.387(96)	& $+86(27)$	& $+53(23)$	& $98(26)$	& $15.9(7.1)$	& $-174(59)$ \\
2014 Jun 08	& 56816.386(96)	& $-7(26)$	& $+20(22)$	& $13(23)$	& $54(52)$	& $-41(57)$ \\
2014 Jun 09	& 56817.385(98)	& $+23(26)$	& $+58(22)$	& $58(22)$	& $34(12)$	& $+63(57)$ \\
2014 Jun 10	& 56818.392(91)	& $+44(27)$	& $+26(23)$	& $46(26)$	& $15(15)$	& $-15(60)$ \\
2014 Jun 11	& 56819.386(98)	& $+20(26)$	& $-29(22)$	& $28(23)$	& $152(24)$	& $-35(57)$ \\
2014 Jun 12	& 56820.387(97)	& $+15(25)$	& $+22(21)$	& $19(22)$	& $28(34)$	& $+39(55)$ \\
2014 Jun 13	& 56821.389(97)	& $-46(25)$	& $+25(22)$	& $47(24)$	& $76(14)$	& $-90(56)$ \\
2014 Jul 10	& 56848.35(14)	& $-20(21)$	& $+11(18)$	& $17(21)$	& $76(32)$	& $-19(47)$ \\
2014 Jul 11	& 56849.35(14)	& $+32(22)$	& $+20(19)$	& $33(21)$	& $16(17)$	& $-109(48)$ \\
2014 Jul 12	& 56850.35(14)	& $+26(21)$	& $+16(18)$	& $25(21)$	& $16(22)$	& $-59(48)$ \\
2014 Jul 13	& 56851.36(14)	& $+46(25)$	& $+65(22)$	& $76(23)$	& $27.3(9.1)$	& $-76(56)$ \\
2014 Jul 14	& 56852.36(14)	& $+44(22)$	& $+27(19)$	& $48(21)$	& $16(12)$	& $+60(49)$ \\
\enddata
\label{189}
\end{deluxetable*}

\begin{deluxetable*}{c c c c c c}
\tabletypesize{\normalsize}
\tablecaption{Binned HD 189733 Observations}
\tablewidth{0pt}
\tablehead{
\colhead{Phase} & \colhead{$q$} & \colhead{$u$} & \colhead{$p$} & \colhead{$\Theta$} & \colhead{$v$} \\
& (ppm) & (ppm) & (ppm) & ($^\circ$) & (ppm)}
\startdata
Time average	& $+15.9(4.0)$	& $+39.5(3.5)$	& $42.4(3.6)$	& $34.1(2.7)$	& $-17.9(9.0)$ \\
\hline
0.061(21)	& $+17(15)$	& $-42(13)$	& $43(13)$	& $145.8(9.7)$	& $-21(33)$ \\
0.150(35)	& $-19(13)$	& $+9(43)$	& $12(22)$	& $-$		& $-7(28)$ \\
0.217(38)	& $-27(13)$	& $+21(11)$	& $32(12)$	& $71(10)$	& $+6(28)$ \\
0.313(29)	& $+15(14)$	& $-6(12)$	& $12(13)$	& $169(28)$	& $-44(30)$ \\
0.407(28)	& $-4(16)$	& $-25(14)$	& $21(14)$	& $130(21)$	& $+35(35)$ \\
0.504(31)	& $+13(11)$	& $+17(49)$	& $13(40)$	& $-$		& $+4(78)$ \\
0.614(23)	& $+1(16)$	& $-13(14)$	& $8(14)$		& $136(49)$	& $+69(35)$ \\
0.697(43)	& $-5(18)$	& $+41(15)$	& $38(15)$	& $48(13)$	& $-17(40)$ \\
0.758(26)	& $+8(14)$	& $-13(12)$	& $11(12)$	& $151(32)$	& $-15(31)$ \\
0.852(27)	& $-22(14)$	& $-22(39)$	& $22(29)$	& $112(41)$	& $+40(30)$ \\
0.958(34)	& $+13(11)$	& $+8.1(9.2)$	& $13(10)$	& $16(21)$	& $+5(24)$ \\
\enddata
\label{189bin}
\end{deluxetable*}

\subsection{Telescope Polarization}

\begin{figure}
\centering
\includegraphics[scale=0.48]{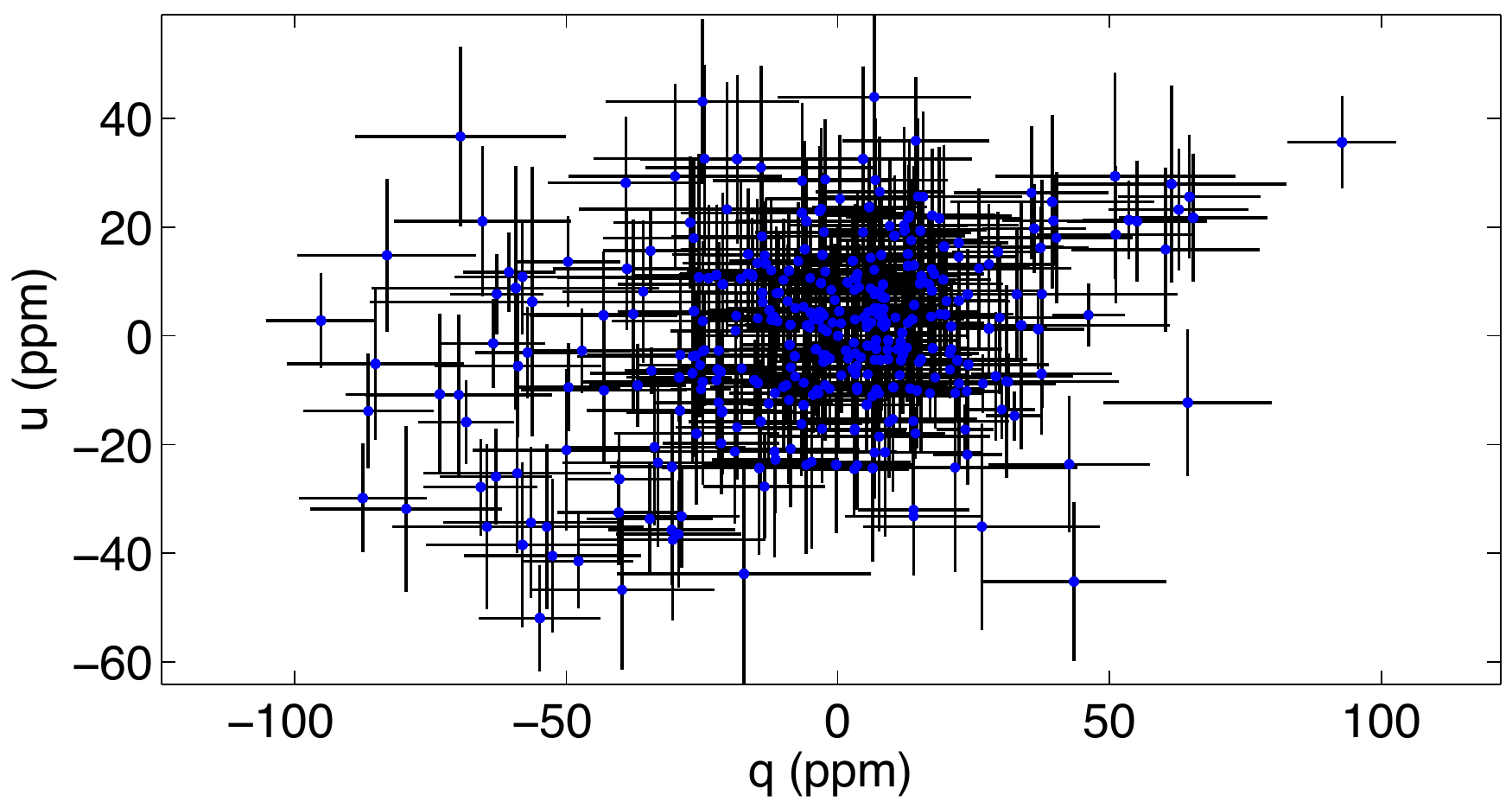}
\caption{Fractional linear polarimetry of all nearly unpolarized calibrator stars.}
\label{tp_qu}
\end{figure}

\begin{figure}
\centering
\includegraphics[scale=0.48]{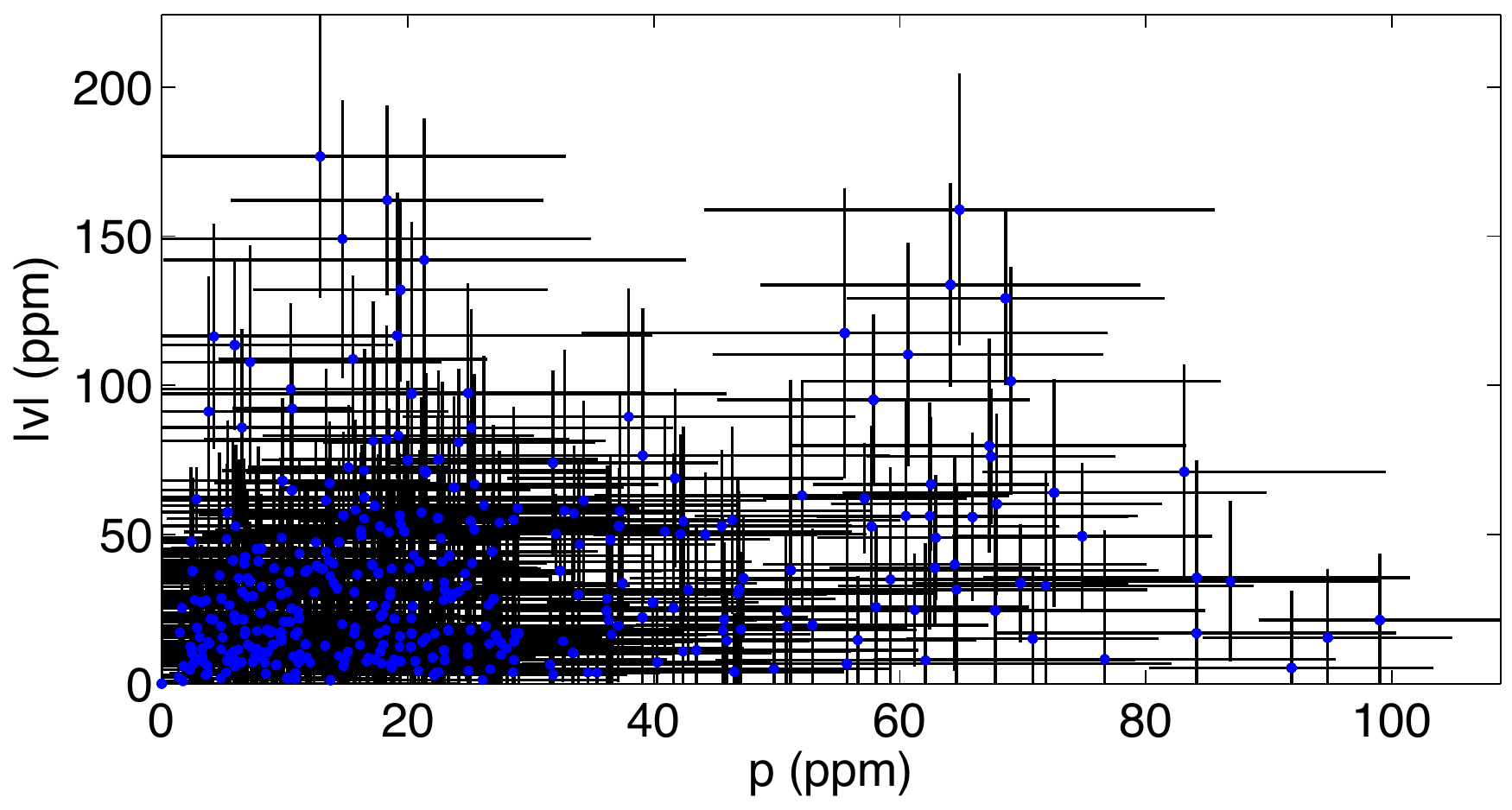}
\caption{Fractional linear and circular polarimetry of all nearly unpolarized calibrator stars.}
\label{tp_pv}
\end{figure}

\begin{figure}
\centering
\includegraphics[scale=0.48]{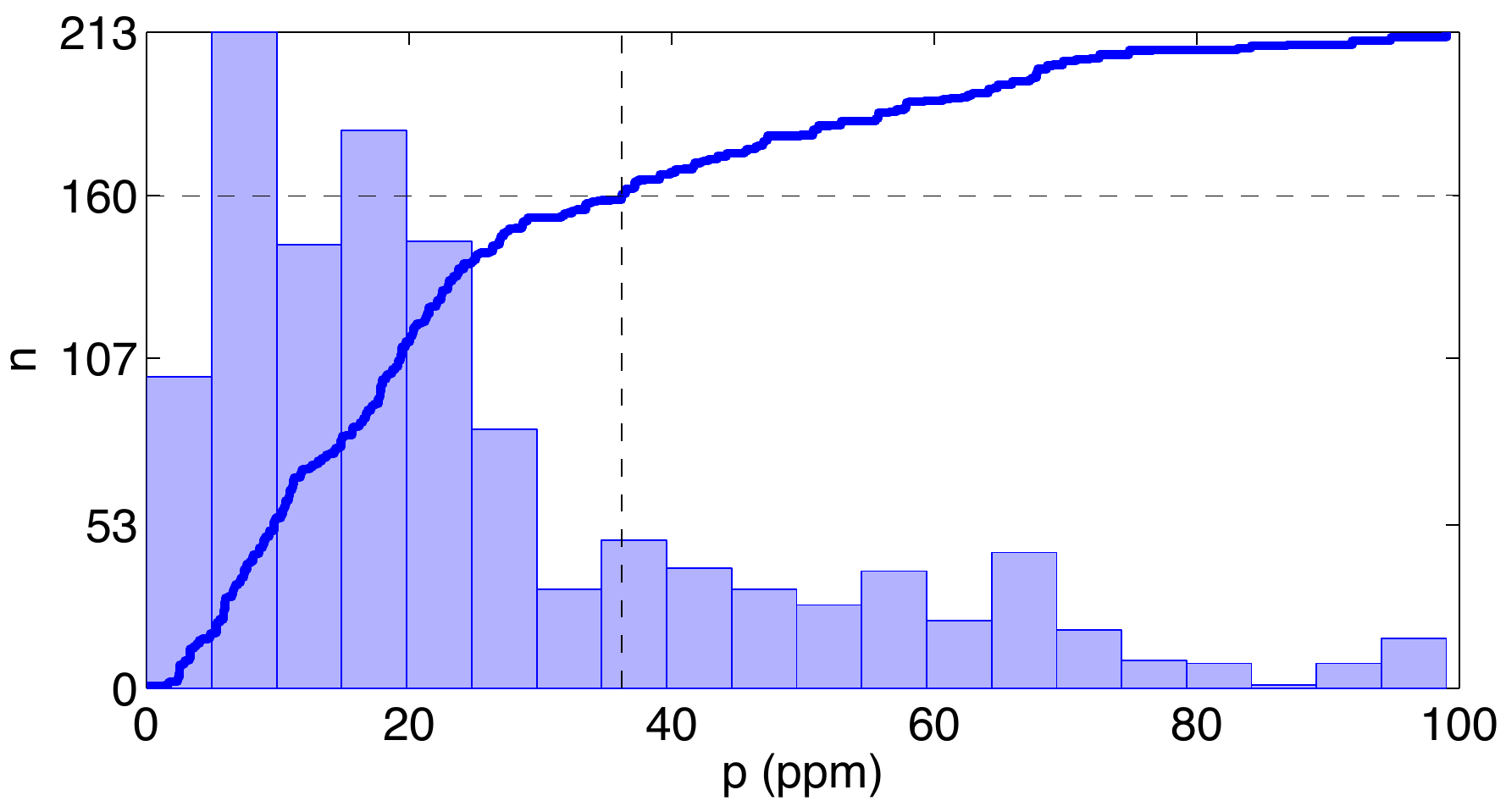}
\caption{Histogram and cumulative distribution function of degree of linear polarization $p$ measurements of nearly unpolarized calibrator stars. 75\% of these measurements have $p < 36$ ppm.}
\label{tp_hist}
\end{figure}

\begin{figure}
\centering
\includegraphics[scale=0.48]{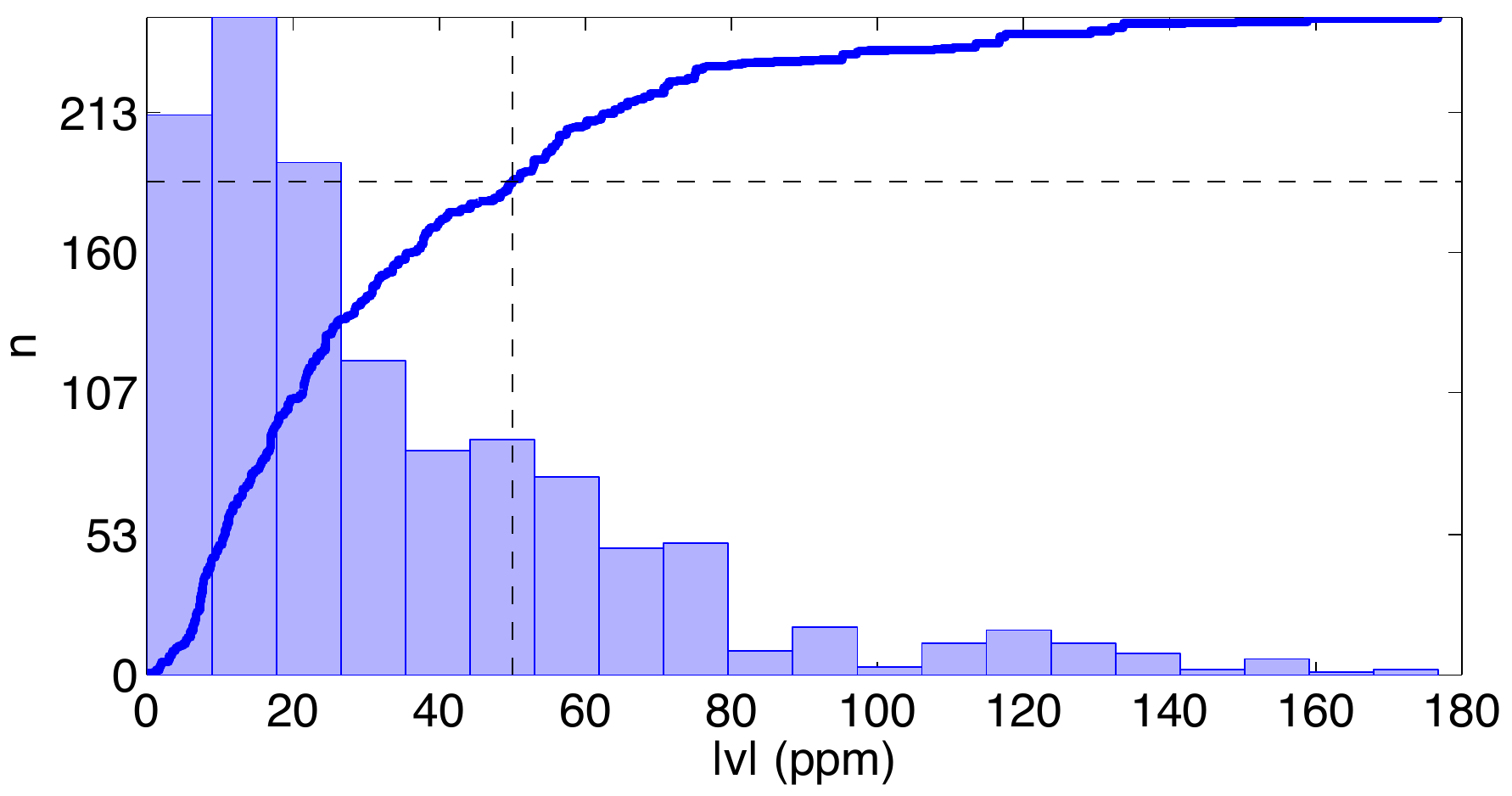}
\caption{Histogram and cumulative distribution function of circular polarization (Stokes $|v|$) measurements of nearly unpolarized calibrator stars. 75\% of these measurements have $p < 50$ ppm.}
\label{tp_histv}
\end{figure}

The 50 nights of HD 189733 observations were obtained over nine observing runs between August 2011 and July 2014. An average of eight nearly unpolarized calibrator stars were observed during each of these runs for a total of 20 individual stars (Table \ref{stars}). A total of 1,229 observations of such calibrator stars were performed, which represents a wealth of information on the $B$ band linear/circular polarization and stability of nearby stars. Indeed, high accuracy observations of nearby stars are essential in understanding the galactic magnetic field in the solar vicinity \citep{Bailey2010, Frisch2010, Frisch2012, Frisch2015}. We determine telescope polarization on a nightly basis from the weighted mean value of each Stokes parameter. Figure \ref{tp_qu} compares linear polarization of these stars in a $q-u$ diagram, where measurements are subtracted by telescope polarization, while linear and circular polarization are compared in Figure \ref{tp_pv}. Both a histogram and the cumulative distribution function of values of $p$ are shown in Figure \ref{tp_hist}. The square root of the weighted variances of all 1,229 observations are $\sigma_q = 26$, $\sigma_u = 13$, and $\sigma_v = 33$ ppm. We find that 75\% of our $B$ band measurements have $p < 36$ ppm, while this value is 20 ppm for the optical red (590 to 1000 nm) observations of \cite{Bailey2010}. Thus, our measurements are broadly consistent with PlanetPol observations of nearby stars \citep{Bailey2010}. We also present a large sample of $B$ band circular polarimetric observations of nearby stars, where 75\% of measurements have $|v| < 50$ ppm (Figure \ref{tp_histv}). This is consistent with the smaller subset of observations published in \cite{WiktorowiczNofi2015}.

%

\subsection{HD 189733}

\begin{figure*}
\centering
\includegraphics[scale=0.75]{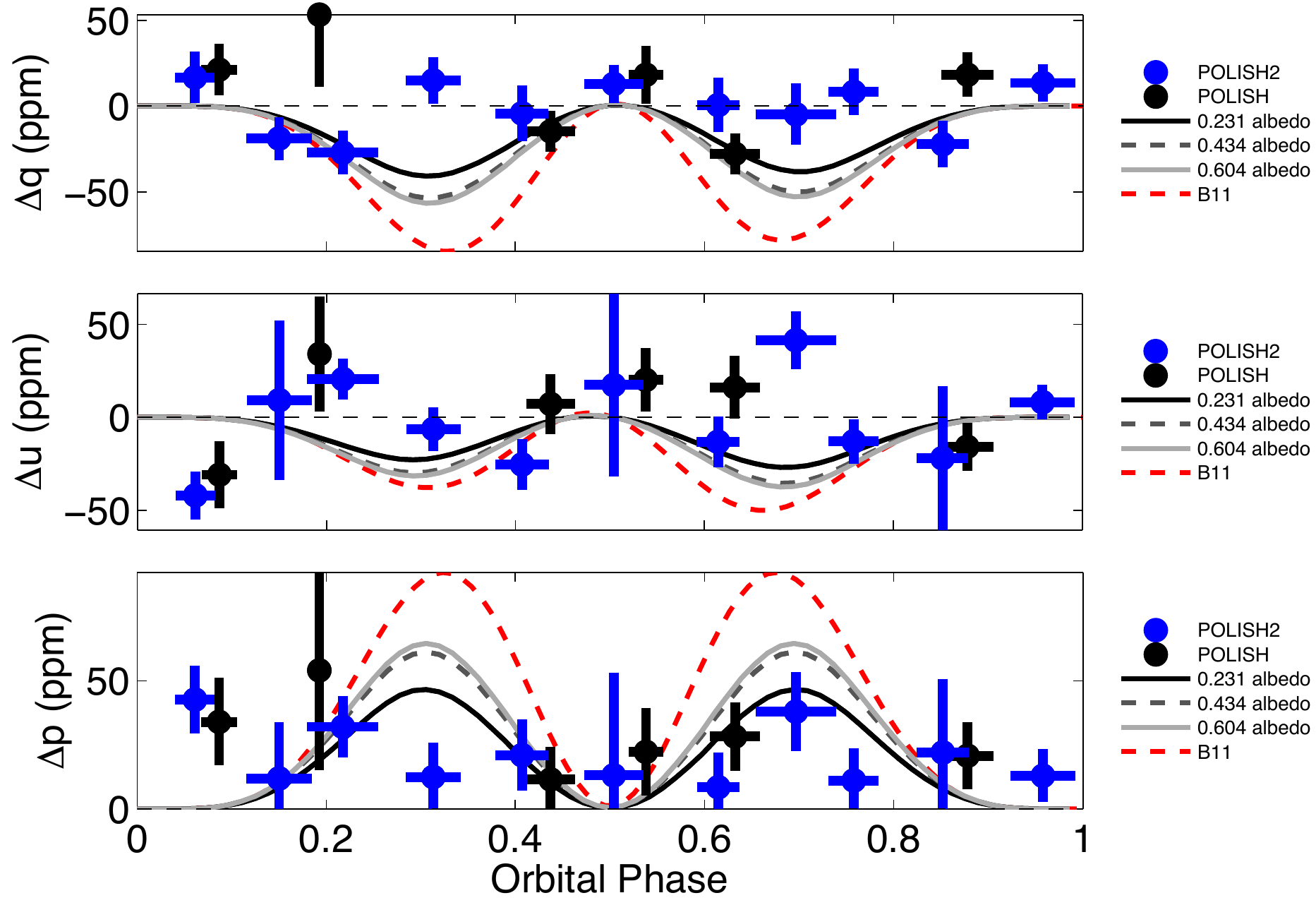}
\caption{Phase-binned observations of HD 189733 versus orbital phase, where phase 0 corresponds to mid-transit and 0.5 to occultation. Lick 3-m/POLISH2 $B$ band data are shown in blue, and Palomar 5-m/POLISH unfiltered data are shown in black \citep{Wiktorowicz2009}. A striking consistency exists between these data, which were taken with two different instruments on two different telescopes. Multiple scattering models with albedos 0.231, 0.434, and 0.604 are shown and compared to a single scattering model with albedo 0.61 \citep[hereafter B11]{Berdyugina2011}. All models use a longitude of the ascending node of $\Omega = 16^\circ$ as from B11.}
\label{189_fig}
\end{figure*}

Figure \ref{189_fig} presents binned Stokes $q$, $u$, and $p$ observations of the HD 189733 system, both from this study (Lick 3-m/POLISH2, $B$ band) and from \cite{Wiktorowicz2009} (Palomar 5-m/POLISH, unfiltered). Both POLISH and POLISH2 measurements are subtracted by their time averages to highlight the time-variable component, and degree of linear polarization $\Delta p$ is recalculated from $\Delta q$ and $\Delta u$. Also shown in Figure \ref{189_fig} are Rayleigh scattering models with geometric albedos of 0.231, 0.434, and 0.604. A detailed description of these models is given in \cite{Kopparla2015}. In addition, we show the model fitting the reported detection from \cite{Berdyugina2011}, which represents a geometric albedo of $0.61 \pm 0.12$ and is calculated using only single scattering. While it is clear that our model cannot be varied to explain our observations, the POLISH and POLISH2 data are strikingly consistent. In particular, a high degree of linear polarization is observed near orbital phase 0.1, which is inconsistent with our Rayleigh scattering model. We do not attempt to investigate the statistical significance or potential cause of this tantalizing observation. We note that while the datasets of \cite{Berdyugina2008} (KVA 0.6-m/DIPol) and \cite{Berdyugina2011} (NOT 2.5-m/TurPol) appear similar, and are also taken with different instrument and telescope combinations, the modest size of these telescopes used does not allow such high accuracy comparisons. \\

It can be seen that statistically significant variations exist from bin to bin in orbital phase, which implies that systematic effects, not photon noise, limit the accuracy of our measurements. On average, five nights of data compose each bin, and the bin uncertainty is given by either the square root of the weighted variance or the standard error of the measurements. This choice is determined by whether the measurements in each bin are inconsistent with each other at the $3 \sigma$ level from a $\chi^2$ test. This tests whether measurements in each bin are drawn from a normally distributed population, because standard error decreases as $1/\sqrt{n}$ only for normally distributed measurements. Indeed, it can be seen that the uncertainties near orbital phase 0.5 and 0.85 are large due to relative disagreement between successive observations at these orbital phases. \\


Even though our measurements are dominated by telescope polarization, the amplitude of the scatter in degree of polarization $p$ is only $\sim 50$ ppm (Figure \ref{189_fig}). This is clearly an interesting accuracy regime, as the scattered light amplitude of HD 189733b has been measured to $\sim 100$ ppm from \textit{HST} \textit{STIS} \citep[hereafter E13]{Evans2013} in total intensity. To determine an upper limit to the polarimetric amplitude due to the planet, and thereby constrain the albedo of the exoplanet, we perform bootstrap resampling of our measurements. For the binned Stokes $q$ and $u$ datasets in Figure \ref{189_fig}, we perform $7,209$ iterations of $10^4$ resampled datasets with the following procedure: 1) the orbital phase value of each bin is randomly assigned from the list of 11 bins, 2) each observable quantity (Stokes $q$ and $u$) is resampled from its nightly mean and uncertainty, and the uncertainty is retained, and 3) degree of polarization $p$ is calculated for each bin from the generalized MAS estimator \citep{Plaszczynski2014}. \\

\begin{figure}
\centering
\includegraphics[scale=0.46]{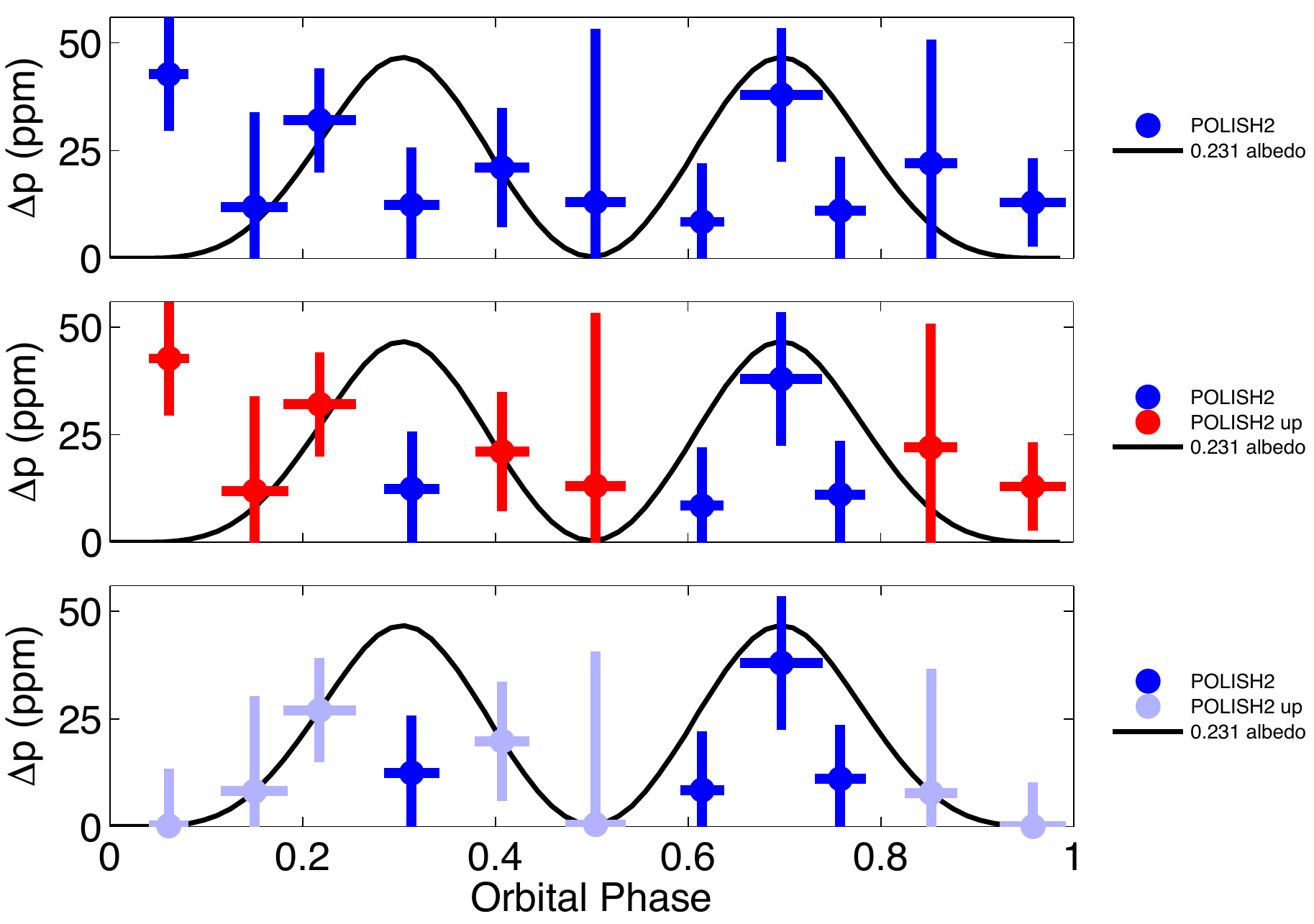}
\caption{\textit{Top:} Observed degree of polarization (Figure \ref{189_fig}, bottom). \textit{Middle:} Measurements in red lie above the model and offer no constraint on albedo. \textit{Bottom:} $\chi^2$ values of measurements lying above the model are set to zero; equivalently, they may be understood as being forced to the model value (lavender points).}
\label{bootproc}
\end{figure}

\begin{figure}
\centering
\includegraphics[scale=0.46]{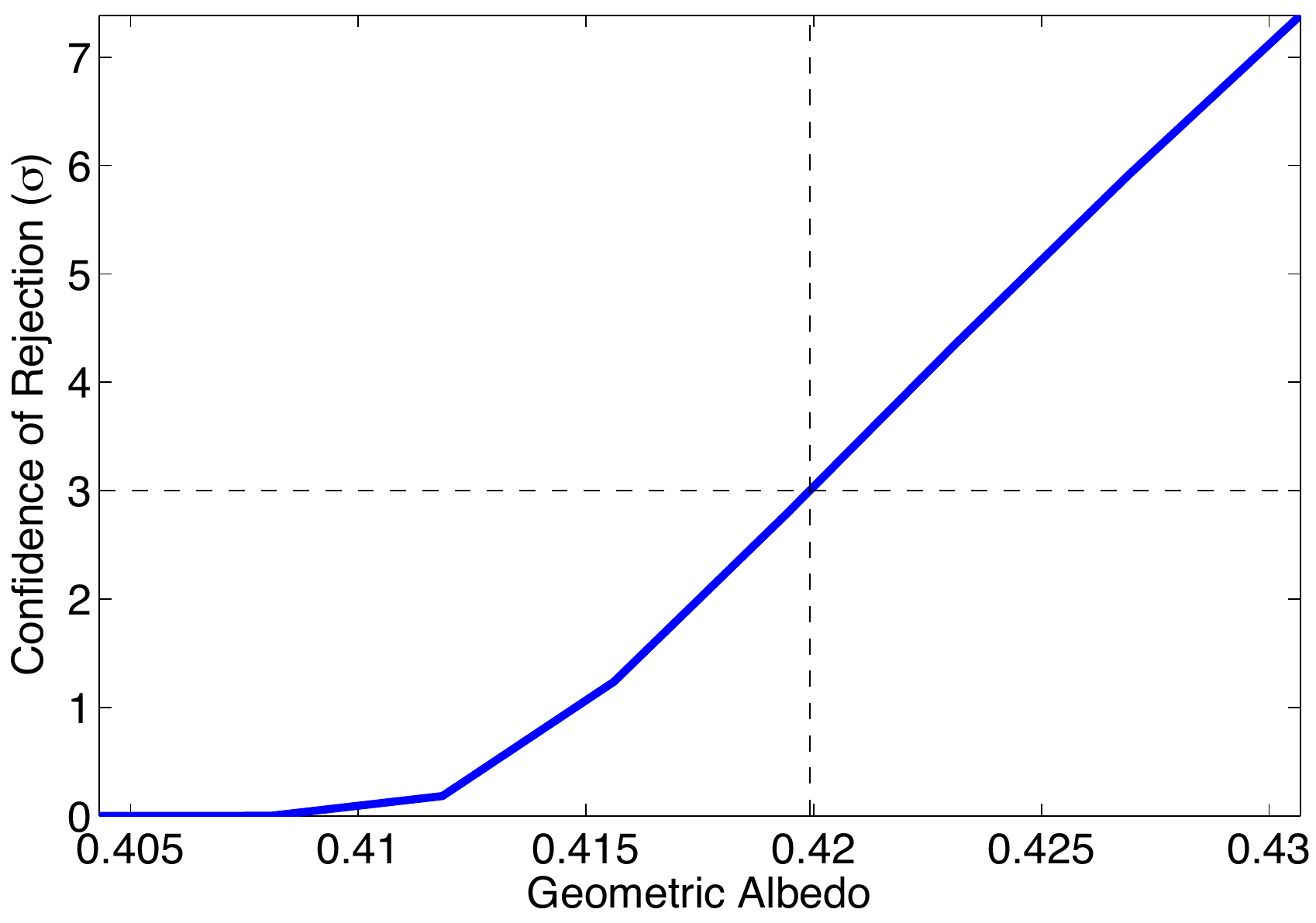}
\caption{Results from a single iteration of $10^4$ resampled datasets, where geometric albedos $> 0.42$ may be rejected with $3 \sigma$ confidence. Over 7,000 additional iterations cause this upper limit to converge with $A_g < 0.37$ (Figure \ref{albedo_rej}).}
\label{bootexample}
\end{figure}

\begin{figure}
\centering
\includegraphics[scale=0.47]{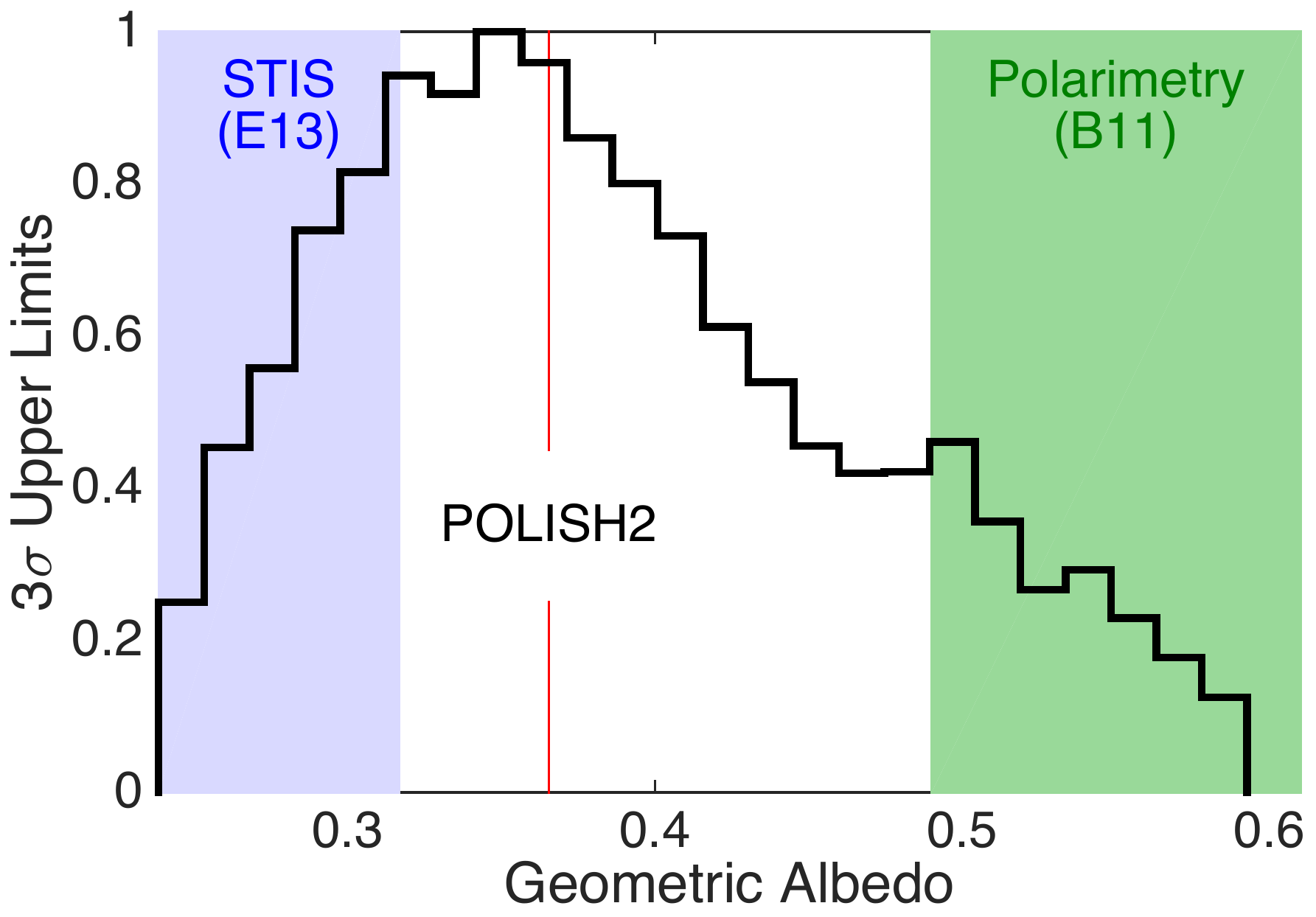}
\caption{Histogram of $3 \sigma$ upper limits to the HD 189733b geometric albedo from $7,209$ iterations of $\chi^2$ analyses between $10^4$ bootstrap resampled datasets (resampled from Figure \ref{189_fig}, bottom panel) and modeled phase curves \citep{Kopparla2015}. Note that a single iteration of $10^4$ resampled datasets is shown in Figure \ref{bootexample}. Geometric albedos larger than 0.37 may be rejected with $> 3 \sigma$ confidence (median of bootstrap histogram, red line). The $0.226 \pm 0.091$ albedo from \textit{HST} \textit{STIS} photometry (E13) cannot be excluded with our data, while the reported $0.61 \pm 0.12$ albedo from polarimetry (B11) may be rejected with $> 99.99\%$ confidence.}
\label{albedo_rej}
\end{figure}

Since Stokes $q$ and $u$ simply represent a rotational transformation from $p$ to the arbitrary observer frame, we perform a $\chi^2$ analysis on the values of $p$ between the bootstrap resamples and the modeled polarization phase curves. However, since our observations are dominated by systematic effects, data with values significantly larger than the models do not rule out those models. Instead, we set positive $\chi^2$ residuals to zero, effectively performing the $\chi^2$ calculation on synthetic data lying below the degree of polarization models (but retaining the degrees of freedom from the full dataset, Figure \ref{bootproc}). The albedo constraint from a single iteration out of $\sim 7 \times 10^3$, which itself contains $10^4$ resampled datasets, is shown in Figure \ref{bootexample}. Figure \ref{albedo_rej} shows the results of this $\chi^2$ analysis, where Rayleigh scattering models with geometric albedos larger than 0.37 may be rejected by our observations with $\geq 3 \sigma$ confidence. By interpolating the amplitude of modeled degree of polarization versus geometric albedo, we find a $3 \sigma$ upper limit to the polarimetric amplitude of HD 189733b to be 58 ppm in $B$ band. \\

\begin{figure}
\centering
\includegraphics[scale=0.41]{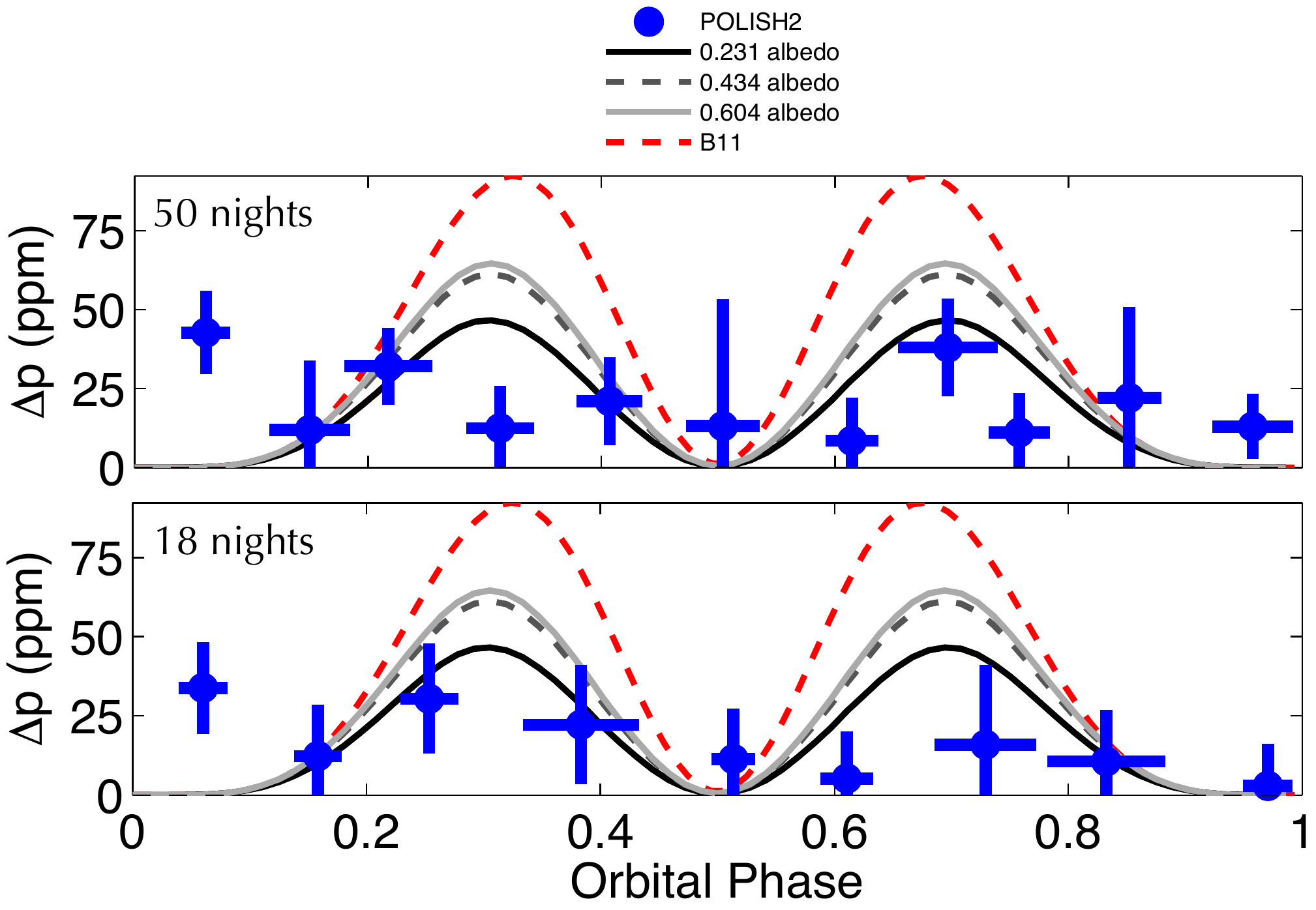}
\caption{Comparison of degree of polarization from the full, 50-night dataset (August 2011 to July 2014) with an 18-night subset (September 2013 to July 2014).}
\label{18nights}
\end{figure}

\section{Discussion}
\label{sec_discuss}

While our dataset is dominated by telescope systematic effects, bootstrap resampling provides a $3 \sigma$ upper limit to the geometric albedo of HD 189733b of $A_g < 0.37$ in $B$ band. The $B$ band POLISH2 bandpass of $391-482$ nm ($\lambda_c = 441$ nm) is essentially identical to the combined $390-435$ nm and $435-480$ nm channels of \textit{STIS}. From \textit{STIS} observations, a weighted mean of the albedo estimates from these channels (E13) suggests an HD 189733b albedo of $0.226 \pm 0.091$ in $B$ band. Figure \ref{albedo_rej} shows that POLISH2 observations are currently unable to constrain the albedo of HD 189733b further than the result of \cite{Evans2013}. \\

However, Figure \ref{albedo_rej} illustrates that even the $1 \sigma$ lower bound to the reported $0.61 \pm 0.12$ albedo from previous polarimetry \citep{Berdyugina2011} is ruled out by POLISH2 observations with over $99.99\%$ confidence. Our current observations, comprising 50 nights of data spanning three years, present a $3 \sigma$ upper limit to the polarimetric modulation of the exoplanet to be 58 ppm in $B$ band. Figure \ref{18nights} shows a subset of our observations spanning 18 nights between September 2013 and July 2014. The similarity between this and the full 50-night dataset (Figure \ref{189_fig}) suggest that accuracy does not continue to improve as $1 / \sqrt{n_{\rm nights}}$, which is expected from systematic effects with non-Gaussian distribution. However, this also shows that high polarimetric accuracy may be obtained from the ground after a reasonable observing campaign. The observations of \cite{Wiktorowicz2009}, taken over six nights in a single run using a different telescope and instrument (Palomar 5-m/POLISH), provide an upper limit of 79 ppm in the $320-633$ nm wavelength range ($\lambda_c = 437$ nm) with 99\% confidence. We note that the wavelength range reported by \cite{Wiktorowicz2009}, roughly $400-675$ nm, neglected a modulation efficiency term; correction for this effect shifts the true bandpass to $320-633$ nm ($\lambda_c = 437$ nm). Therefore, we confirm (and extend to $B$ band) the conclusion of \cite{Wiktorowicz2009} that the large, $\sim 100$ ppm polarimetric amplitude reported from previous polarimetry \citep{Berdyugina2008, Berdyugina2011} cannot be due to polarized, scattered light from the HD 189733b hot Jupiter. \\

Indeed, given the known radius of the exoplanet from transit photometry, a large, $A_g \approx 0.6$ albedo requires a model including single scattering only. This is because the inclusion of multiple scattering, expected in an atmosphere with high albedo, acts to depolarize light scattered by the exoplanet atmosphere \citep{Buenzli2009, Lucas2009, Kopparla2015}. However, the long-standing observation of circular polarization in scattered light from the poles of Jupiter \citep{Kemp1971a, Kemp1971b, Michalsky1974} requires the presence of multiple scattering. Thus, it is reasonable to expect that since multiple scattering is required to understand the scattered light from Jupiter, with geometric albedo $A_g = 0.52$, models of scattered light from Jovian exoplanets must also include multiple scattering. \\

We suggest that hitherto unidentified systematic effects cause spurious polarization measurements with the polarimeters used in previously reported exoplanet polarimetry \citep{Berdyugina2008, Berdyugina2011}. While PEM polarimeters such as PlanetPol and POLISH2 have demonstrated accuracies of order 10 ppm in the literature on inter- \citep{Bailey2010} and intra-night timescales \citep{WiktorowiczNofi2015}, we suggest that the slow modulation and asynchronous observation of Stokes $q$ and $u$ inherent in waveplates may impede scattered light detections with conventional polarimeters. \\

Our difficulty in the accurate correction for telescope polarization lies with the equatorial mount of the Lick 3-m telescope. This is because stellar and telescope polarization cannot be separated, such as by parallactic angle rotation on alt-az telescopes. This requires nightly observation of nearly unpolarized stars, whose weak polarization is a direct result of their proximity to Earth due to the linear dependence of interstellar polarization with heliocentric distance (e.g., \citealp{Hall1949, Hiltner1949, Fosalba2002}). Since these calibration stars are by definition nearby and bright, they are necessarily distributed nearly uniformly across the sky. Thus, it is possible that the varying gravitational environment of the telescope introduces a change to the telescope polarization between science and calibration targets. The addition of data at similar exoplanet orbital phases, but taken during different times of year, may include a variable component of telescope polarization. However, self calibration of telescope polarization, during observations of the science target, has been shown to enable high accuracy for the Gemini Planet Imager \citep{Wiktorowicz2014}, which is mounted at the alt-az Gemini South 8-m telescope. Therefore, it is expected that a POLISH2-like instrument mounted at an alt-az telescope of greater than 3-m aperture will provide the highest accuracy observations necessary to detect scattered light from close-in exoplanets. This represents the best of both worlds, utilizing the superior accuracy of photoelastic modulators over waveplates and the enhanced calibration environment of alt-az telescopes over equatorial ones. \\

However, even from a 60-year-old, equatorial 3-m telescope overlooking the tenth largest city in the US, we have demonstrated accuracy sufficient to test specific exoplanet results from the space-based, 2.4-m Hubble Space Telescope. From our extensive observations of nearly unpolarized calibrator stars, we plan to correlate telescope polarization with telescope altitude and azimuth to enable a flexure correction to Lick 3-m POLISH2 observations. In addition, beginning with our June 2014 observations, we have altered our calibration strategy to bracket exoplanet observations with periodic observations of the same nearly unpolarized calibrator stars during each night. These observations will further improve the accuracy of flexure correction. Finally, rather than binning mean nightly data in orbital phase, we will re-bin data from run to run according to their 0.1 sec data segments. This enables any variability on hourly timescales to be binned properly, which may reduce systematic effects and increase accuracy. \\

Even absent these improvements to calibration, POLISH2 accuracy is currently capable of detecting polarization of any close-in exoplanet with an amplitude larger than 58 ppm given sufficient observing time. While HD 189733b was observed due to the combination of 1) close-in orbit, 2) large radius, and 3) bright host star, other targets exist whose expected polarimetric amplitudes are larger than our systematic noise floor. We have already begun observations of such exoplanets. \\

\section{Conclusion}
\label{section_conclusion}

Using 50 nights of HD 189733 observations with the POLISH2 polarimeter at the Lick 3-m telescope, we constrain the geometric albedo of HD 189733b to be $A_g < 0.37$ in the $B$ band with $3 \sigma$ confidence. This value is consistent with the $0.226 \pm 0.091$ albedo from \textit{STIS} photometry determined by \cite{Evans2013}, but we reject the reported $0.61 \pm 0.12$ albedo from previous polarimetry with over $99.99\%$ confidence \citep{Berdyugina2011}. The conclusive detection of Rayleigh scattering from this exoplanet is of the highest significance for exoplanet science, because HD 189733b has become the poster child for the recent finding of hazes on exoplanets \citep{Lecavelier2008, Pont2008, Sing2009, Sing2011, Gibson2012, Huitson2012, Pont2013}. However, photometric techniques potentially have a mundane explanation for tantalizing suggestions of Rayleigh scattering, as inaccurate subtraction of unocculted starspots may masquerade as a signature of Rayleigh scattering \citep{McCullough2014}. Given the requirement that high accuracy exoplanet science be repeatable, and given the predisposition to resort to space-based inquiry, we demonstrate the virtue of long temporal baseline, ground-based study of exoplanets using complementary techniques afforded by the physics of the photon. \\

While our observations are limited by systematic effects at the 58 ppm level, which are inherent in equatorial telescopes, we are currently observing a sample of exoplanets expected to provide larger amplitudes in polarized light. We are also in the process of performing additional calibration measures, via empirical telescope flexure correction, that may reduce the systematic noise floor from the Lick 3-m telescope. Additionally, we have previously demonstrated the utility of self-calibration on science targets with polarimeters at large, alt-az telescopes \citep{Wiktorowicz2014}. Therefore, we advocate for POLISH2-like, photoelastic polarimeters at large, modern telescopes for the conclusive detection of scattered light from a large sample of close-in exoplanets. \\

\acknowledgments
We would like to acknowledge the tireless efforts of the Lick Observatory staff. This work was performed (in part) under contract with the California Institute of Technology (Caltech) funded by NASA through the Sagan Fellowship Program executed by the NASA Exoplanet Science Institute. SJW and LAN acknowledge support from the NASA Origins of Solar Systems program through grant NNX13AF63G. PK and YLY acknowledge support from an NAI Virtual Planetary Laboratory grant from the University of Washington to the Jet Propulsion Laboratory and California Institute of Technology. Research at Lick Observatory is partially supported by a generous gift from Google. \\
 
 {\it Facilities:} \facility{Shane (POLISH2)}.


 \end{document}